\documentclass{article}

\usepackage{microtype}
\usepackage{graphicx}
\usepackage{subfigure}
\usepackage{amsfonts}
\usepackage{booktabs} 
\usepackage{amsmath}
\usepackage{amsthm}
\usepackage{wrapfig}
\usepackage{tikz}

\usepackage{hyperref}

\newtheorem{definition}{Definition}
\newtheorem{lemma}{Lemma}
\newtheorem{proposition}{Proposition}

\newcommand*\circled[1]{\tikz[baseline=(char.base)]{
            \node[shape=circle,draw,inner sep=2pt] (char) {#1};}}
\makeatletter
\newcommand{\distas}[1]{\mathbin{\overset{#1}{\kern\z@\sim}}}%

\usepackage[accepted]{icml2021}

\icmltitlerunning{Information-Guided Sampling for Low-Rank Matrix Completion}

\begin{document}

\twocolumn[
\icmltitle{Information-Guided Sampling for Low-Rank Matrix Completion}



\icmlsetsymbol{equal}{*}

\begin{icmlauthorlist}
\icmlauthor{Simon Mak}{duke}
\icmlauthor{Henry Shaowu Yuchi}{gt}
\icmlauthor{Yao Xie}{gt}
\end{icmlauthorlist}

\icmlaffiliation{gt}{H. Milton Stewart School of Industrial and Systems Engineering, Georgia Institute of Technology, Georgia, USA}
\icmlaffiliation{duke}{Department of Statistical Science, Duke University, North Carolina, USA}

\icmlcorrespondingauthor{Henry Shaowu Yuchi}{shaowu.yuchi@gatech.edu}

\icmlkeywords{Machine Learning, ICML}

\vskip 0.3in
]




\begin{abstract}
The matrix completion problem, which aims to recover a low-rank matrix $\mathbf{X}$ from partial, noisy observations of its entries, arises in many machine learning applications. In this work, we present a novel information-theoretic framework for initial and sequential sampling of matrix entries for noisy matrix completion, based on the maximum entropy sampling principle in \citet{SW1987}. The key novelty in our approach is that it makes use of uncertainty quantification (UQ) -- a measure of uncertainty for unobserved entries -- to guide the sampling procedure. Our framework reveals new insights on the role of coherence and coding design on sampling for matrix completion. 
\end{abstract}
\vspace{-0.2in}
\section{Introduction}
Low-rank matrices play a vital role in modeling many scientific and engineering problems, including (but not limited to) image processing, satellite imaging, and network analysis. In such applications, however, only a small portion of the desired matrix (which we denote as $\mathbf{X} \in \mathbb{R}^{m_1 \times m_2}$ in this article) can be observed. The reasons for this are two-fold: (i) the cost of observing all matrix entries can be high 
(e.g., in gene studies \cite{ND2014}); (ii) there can be missing observations at individual entries.
The \textit{matrix completion} problem aims to complete the missing entries of $\mathbf{X}$ from partial (and often noisy) observations. 

Matrix completion has attracted much attention since the seminal works of \citet{CT2010}, \citet{CR2012}, and \citet{Rec2011}. This is then extended to the \textit{noisy} matrix completion setting, where entries are observed with noise; important results include \citet{CP2010}, \citet{Rea2010}, \citet{Kea2011}, and \citet{NW2012}, among others. There is now a rich body of work on matrix completion with various convex and non-convex optimization algorithms; recent overviews include \citet{DR2016} and \citet{chi2019nonconvex}. However, much of the literature has focused on the \textit{point estimation} of $\mathbf{X}$ only, with little work on quantifying the uncertainty of such estimates and how this can help guide the sampling of matrix entries. 

This work presents a novel approach for \textit{designing} the observed entries in $\mathbf{X}$ for low-rank matrix completion, with the goal of maximizing information on $\mathbf{X}$. While most of the literature assumes entries are sampled uniformly at random, there has been some work on adaptive sampling schemes \cite{Sea2013, Lea2016, Bea2017, Rea2015}. There is also related work from binary relation learning \cite{GW1995,NA2002}, but such methods do not consider low-rank structure. 
This paper makes novel contributions to the topic of \textit{information-theoretic} sampling for low-rank matrix completion. Our work differs from the large body of literature on information-theoretic design \cite{PV2006, Cea2012, Wea2014, shlezinger2017measurement} in that, instead of maximizing the mutual information between signal (i.e., $\mathbf{X}$) and observed entries (denoted as $\mathbf{Y}_{\Omega}$), we study a dual but equivalent problem of maximizing the \textit{entropy} of observations $\mathbf{Y}_{\Omega}$. Using the maximum entropy sampling principle from \cite{SW1987}, this dual view sheds new insights into matrix completion sampling and provides simple closed-form criteria for initial and active learning.
\vspace{-0.1in}
\section{Low Rank Matrix Modeling}
\label{sec:mod}


\subsection{The completion problem}
Let $\mathbf{X} = (X_{i,j}) \in \mathbb{R}^{m_1 \times m_2}$ be the desired low-rank matrix. Suppose $\mathbf{X}$ is observed with noise at $N$ indices $\Omega_{1:N} = \{(i_n,j_n)\}_{n=1}^N \subseteq [m_1] \times [m_2]$\footnote{$[m] := \{1, \cdots, m\}$, $m_1 \wedge m_2 := \min(m_1,m_2)$, $m_1 \vee m_2 := \max(m_1, m_2)$. \label{fn1}} (sometimes denoted as $\Omega$ for brevity). Let $Y_{i,j}$ be the noisy observation at index $(i,j) \in \Omega$ under the Gaussian noise model:
\begin{equation}
Y_{i,j} = X_{i,j} + \epsilon_{i,j}, \quad \epsilon_{i,j} \distas{i.i.d.} \mathcal{N}(0,\eta^2).
\label{eq:mc}
\end{equation}
Further let $\mathbf{X}_{\Omega} \in \mathbb{R}^N$ and $\mathbf{Y}_{\Omega} \in \mathbb{R}^N$ denote the vectorized entries of $\mathbf{X}$ and $\mathbf{Y}$ at observed indices $\Omega$, and let $\mathbf{X}_{\Omega^c} \in \mathbb{R}^{m_1 m_2 - N}$ and $\mathbf{Y}_{\Omega^c} \in \mathbb{R}^{m_1 m_2 - N}$ denote the vectorized entries of $\mathbf{X}$ and $\mathbf{Y}$ at unobserved indices $\Omega^c = ([m_1] \times [m_2]) \setminus \Omega$.


\subsection{Singular Matrix-variate Gaussian model}
\label{sec:spec}

Consider the following model for the low-rank matrix $\mathbf{X}$ (assumed to be normalized with zero mean):
\begin{definition}
[Singular matrix-variate Gaussian (SMG); Definition 2.4.1, \cite{GN1999}] Let $\mathbf{Z} \in \mathbb{R}^{m_1 \times m_2}$ be a random matrix with entries $Z_{i,j} \distas{i.i.d.} \mathcal{N}(0, \sigma^2)$ for $(i,j) \in [m_1] \times [m_2]$. The random matrix $\mathbf{X}$ has a \textup{singular matrix-variate Gaussian (SMG)} distribution if $\mathbf{X} \stackrel{d}{=} \mathcal{P}_{\mathcal{U}} \mathbf{Z} \mathcal{P}_{\mathcal{V}}$ for some projection matrices $\mathcal{P}_{\mathcal{U}} = \mathbf{U}\mathbf{U}^T$ and $\mathcal{P}_{\mathcal{V}} = \mathbf{V}\mathbf{V}^T$, where $\mathbf{U} \in \mathbb{R}^{m_1 \times R}$, $\mathbf{U}^T\mathbf{U} = \mathbf{I}$,  $\mathbf{V} \in \mathbb{R}^{m_2 \times R}$, $\mathbf{V}^T\mathbf{V} = \mathbf{I}$ and $R < m_1 \wedge m_2 {}^{\ref{fn1}}$. 
\label{def:smg}
\end{definition}
\vspace{-0.1in} 
The projection matrices $\mathcal{P}_{\mathcal{U}}$ and $\mathcal{P}_{\mathcal{V}}$ provide a parametrization of the row and column subspaces of $\mathbf{X}$. Since such parameters are unknown in practice, we adopt a Bayesian approach \cite{Gea2014} and assign the following non-informative priors as in \citet{yuchi2021bayesian}:
\begin{align}
\begin{split}
&[\mathcal{P}_{\mathcal{U}}] \sim \text{Unif}(\mathcal{G}_{R,m_1-R}),\quad [\mathcal{P}_{\mathcal{V}}] \sim \text{Unif}(\mathcal{G}_{R,m_2-R}),\\
&[\eta^2] \sim IG(\alpha_{\eta^2},\beta_{\eta^2}), \quad [\sigma^2] \sim IG(\alpha_{\sigma^2}, \beta_{\sigma^2})
\label{eq:prior}
\end{split}
\end{align}
where $\mathcal{G}_{R,m-R}$ is the \textit{Grassmann manifold} \cite{Chi2012} containing all $m \times m$ projection matrices of rank $R$.

Under such priors, the maximum-a-posteriori (MAP) estimator of matrix $\mathbf{X}$ is closely connected to the nuclear-norm formulation, widely used for matrix completion \cite{CR2012,CT2010}: 
\vspace{-0.05in}
\begin{equation}
\small
\hat{\mathbf{X}} = \underset{\mathbf{X} \in \mathbb{R}^{m_1 \times m_2}}{\textup{Argmin}} \; \Big[ \sum_{(i,j) \in \Omega} (Y_{i,j} - X_{i,j})^2 + \lambda \|\mathbf{X}\|_* \Big] .
\label{eq:nuc}
\vspace{-0.05in}
\end{equation}
Here, $\|\mathbf{X}\|_*$ is the nuclear norm which sums up the singular values of $\mathbf{X}$.
Further details on this connection can be found in \citet{yuchi2021bayesian}.

\vspace{-0.1in}
\section{Maximum entropy sampling for matrix completion}
\label{sec:des}

With this modeling framework, we present a information-theoretic method for sampling matrix entries, using the \textit{maximum entropy principle}, which was first introduced in \citet{SW1987} for experimental design of spatio-temporal models. This has two parts: (a) an \textit{initial sampling} strategy for preliminary learning on $\mathbf{X}$, and (b) a \textit{sequential sampling} strategy to greedily maximize information gain. 

\subsection{The maximum entropy sampling principle}
In the following, we use the definitions of entropy as presented in \citet{CT2012}. For two random variables $(X,Y)$, the chain rule (Theorem 2.2.1, \citet{CT2012}) connects the \textit{joint entropy} ${\rm H}(X,Y)$ with the \textit{conditional entropy} ${\rm H}(Y|X)$ via the identity ${\rm H}(X,Y) = {\rm H}(X) + {\rm H}(Y|X)$.
Consider now the noisy matrix completion problem. Applying this chain rule, we get the decomposition:
\begin{equation}
{\rm H}(\mathbf{Y}_{\Omega},\mathbf{X}) = {\rm H}(\mathbf{Y}_{\Omega}) + {\rm H}(\mathbf{X}|\mathbf{Y}_{\Omega}).
\label{eq:chmat}
\end{equation}
Here, ${\rm H}(\mathbf{Y}_{\Omega},\mathbf{X})$ is the joint entropy of observations $\mathbf{Y}_{\Omega}$ and matrix $\mathbf{X}$, ${\rm H}(\mathbf{Y}_{\Omega})$ is the entropy of $\mathbf{Y}_{\Omega}$, and ${\rm H}(\mathbf{X}|\mathbf{Y}_\Omega)$ is the conditional entropy of $\mathbf{X}$ after observing $\mathbf{Y}_{\Omega}$. To \textit{maximize} information gain on $\mathbf{X}$ from observations, we wish to sample indices $\Omega$ which \textit{minimize} conditional entropy ${\rm H}(\mathbf{X}|\mathbf{Y}_{\Omega})$.


Next, for the sampling model \eqref{eq:mc}, it can be shown that the joint entropy ${\rm H}(\mathbf{Y}_{\Omega},\mathbf{X})$ does \textit{not} depend on sampled indices $\Omega$ (this follows by reversing $\mathbf{X}$ and $\mathbf{Y}_{\Omega}$ in \eqref{eq:chmat}, see \citet{MX2017} for details). Hence, a sampling scheme $\Omega$ which maximizes the entropy of \textit{observations} $\mathbf{Y}_{\Omega}$ (i.e., ${\rm H}(\mathbf{Y}_{\Omega})$) also minimizes the conditional entropy ${\rm H}(\mathbf{X}|\mathbf{Y}_{\Omega})$, which yields maximum \textit{information gain} on $\mathbf{X}$. The maximum entropy sampling principle for noisy matrix completion therefore aims to maximize ${\rm H}(\mathbf{Y}_{\Omega})$. The key advantage of this principle is that it allows us to work with a simple, closed-form expression for ${\rm H}(\mathbf{Y}_{\Omega})$ (derived below) as an efficient proxy for the desired entropy term ${\rm H}(\mathbf{X}|\mathbf{Y}_\Omega)$, which is more complicated and difficult to optimize.

From the SMG model, the exp-entropy\footnote{Since $\exp(\cdot)$ is monotone, the maximum entropy principle holds when maximizing exp-entropy ${\rm E}({\Omega}_{1:N})$.} term ${\rm E}({\Omega}_{1:N}) := \text{exp}\{{\rm H}(\mathbf{Y}_{\Omega}) \}$ admits a closed form:\\
\begin{lemma}[Observational entropy] For fixed $\mathcal{P}_{\mathcal{U}}$ and $\mathcal{P}_{\mathcal{V}}$,
\small
\begin{equation}
{\rm E}({\Omega}_{1:N}) := \exp\{ {\rm H}(\mathbf{Y}_{\Omega}) \} = C \textup{det}\{ \sigma^2 \mathbf{R}_N(\Omega_{1:N}) + \eta^2 \mathbf{I}\},
\label{eq:obsent}
\end{equation}
\normalsize
for some constant $C$ not depending on indices $\Omega_{1:N}$.
\label{lem:obsent}
\end{lemma}
Our strategy is as follows. For \textit{initial} sampling (Section \ref{sec:ini}), we will first derive a lower bound on \eqref{eq:obsent}, then generalize this bound under uniform (non-informative) priors on $(\mathcal{P}_{\mathcal{U}}, \mathcal{P}_{\mathcal{V}})$. The maximization of this generalized bound yields an intuitive initial sampling strategy, with connections to coding design. For \textit{sequential} sampling (Section \ref{sec:seq}), we will first use the nuclear-norm minimization in \eqref{eq:nuc} to obtain subspace estimates, then plug in these estimates into \eqref{eq:obsent} to derive an efficient, sequential sampling algorithm. 

\subsection{Initial sampling: Latin square design}
\label{sec:ini}
For simplicity, assume $m_1 = m_2 = m$ (generalized later), with total initial samples $N = m$. Following \citet{yuchi2021bayesian}, we define the \textit{coherence} of subspace $\mathcal{U}$ for the $i$-th basis vector $\mathbf{e}_i$ as $\mu_i(\mathcal{U}) := \|\mathcal{P}_{\mathcal{U}} \mathbf{e}_i\|_2^2$, and the \textit{cross-coherence} of $\mathcal{U}$ for bases $\mathbf{e}_i$ and $\mathbf{e}_{i'}$ as $\nu_{i,i'}(\mathcal{U}) = \mathbf{e}_{i'}^T \mathcal{P}_{\mathcal{U}}\mathbf{e}_{i}$. The lemma below gives a lower bound on ${\rm E}(\Omega_{1:N})$:
\begin{proposition}[Lower bound on observation entropy]
For fixed $\mathcal{P}_{\mathcal{U}}$ and $\mathcal{P}_{\mathcal{V}}$, we have
\begin{equation}
\small
\begin{split}
\small
&{\rm E}^{1/N}(\Omega_{1:N}) \geq  \min_{n=1, \cdots, N} C \Bigg[ \sigma^2 \mu_{i_n}(\mathcal{U}) \mu_{j_n}(\mathcal{V}) + \eta^2 - \Psi(\Omega_{1:N}) \Bigg],
\label{eq:ev}
\end{split}
\end{equation}
where:
\begin{equation}
\small
\Psi(\Omega_{1:N}) = \frac{\sigma^2 (N-1)}{2R} \left\{ \max_{n':n' \neq n} \nu_{i_n,i_{n'}}^2(\mathcal{U}) + \max_{n':n' \neq n} \nu_{j_n,j_{n'}}^2(\mathcal{V}) \right\}
\label{eq:ev2}
\end{equation}
\normalsize
and $C$ is a constant not depending on $\Omega_{1:N}$.
\label{prop:balance}
\normalsize
\end{proposition}


Using the right side of \eqref{eq:ev} as a proxy for ${\rm E}^{1/N}(\Omega_{1:N})$, the maximization of ${\rm E}(\Omega_{1:N})$ under the non-informative priors \eqref{eq:prior} can be approximated by the minimization of \eqref{eq:ev2}, since the coherence terms $\mu_{i_n}(\mathcal{U})$ and $\mu_{j_n}(\mathcal{V})$ are on average equal for any $i_n$ and $j_n$, under uniform priors on $\mathcal{P}_{\mathcal{U}}$ and $\mathcal{P}_{\mathcal{V}}$. This minimization amounts to jointly minimizing $\max_{n \neq n'} (\mathbf{e}_{i_n}^T\mathbf{e}_{i_{n'}})^2$ and $\max_{n \neq n'} (\mathbf{e}_{j_n}^T\mathbf{e}_{j_{n'}})^2$. Clearly, if $i_n = i_{n'}$ for some $n \neq n'$ (i.e., same row is sampled twice), then the first term attains the maximum value of 1. Likewise, if $j_n = j_{n'}$ for some $n \neq n'$ (i.e., same column is sampled twice), then the second term becomes 1 as well. Both are undesirable, since we wish to minimize the two terms in \eqref{eq:ev2}. 

With little knowledge on $\mathbf{X}$, an initial sampling scheme achieving maximum entropy should therefore be \textit{balanced}, in that all rows and columns should be sampled exactly once (in specific case of $N=m_1=m_2$) or an equal number of times (for general matrices).
This balanced sampling scheme can be nicely represented by a {\it Latin square} \cite{keedwell2015latin}: an $m \times m$ array with $m$ distinct symbols, each occurring exactly once in each row and column. Figure \ref{fig:latin} shows an example of a $3 \times 3$ and $4 \times 4$ Latin square; a balanced sample is obtained by choosing indices with a specific letter (e.g., `a'). Latin squares are widely used in error-correcting codes \cite{Cea2004, Huc2006} and experimental design \cite{Fis1937}, and there are efficient algorithms \cite{JM1996} for generating randomized Latin squares.

\begin{figure}[t]
\small
\[ \begin{pmatrix}
\circled{\textit{a}} & c & b\\
c & b & \circled{\textit{a}}\\
b & \circled{\textit{a}} & c\\
\end{pmatrix} \quad  \quad \begin{pmatrix}
\circled{\textit{a}} & b & c & d\\
c & d & \circled{\textit{a}} & b\\
d & c & b & \circled{\textit{a}}\\
b & \circled{\textit{a}} & d & c\\
\end{pmatrix} \]
\normalsize
\vspace{-0.5cm}
\caption{A $3 \times 3$ and a $4 \times 4$ Latin square. A balanced sampling scheme is obtained by sampling the entries with `\textit{a}' (circled).}
\label{fig:latin}
\end{figure}

\subsection{Sequential sampling}
\label{sec:seq}
Consider next the setting where the noisy entries $\mathbf{Y}_{\Omega}$ have been observed at $\Omega_{1:N}$, and suppose informed estimates are on subspaces $\mathcal{U}$ and $\mathcal{V}$ from $\mathbf{Y}_{\Omega}$ (more on this in Section \ref{sec:me}). Fixing the observed indices $\Omega_{1:N}$, the sequential problem of sampling the next index $(i,j) \notin \Omega_{1:N}$ maximizing exp-entropy ${\rm E}(\Omega_{1:N} \cup (i,j))$ can be formulated as:
\begin{lemma}
For fixed $\mathcal{P}_{\mathcal{U}}$, $\mathcal{P}_{\mathcal{V}}$ and observed indices $\Omega_{1:N}$,
\small
\begin{align}
\small
\begin{split}
&\underset{(i,j) \in \Omega_{1:N}^c}{\textup{Argmax}} \; \textup{E}(\Omega_{1:N} \cup (i,j)) \\
& = \underset{(i,j) \in \Omega_{1:N}^c}{\textup{Argmax}} \; \{ \mu_i(\mathcal{U}) \mu_j(\mathcal{V}) 
-\boldsymbol{\nu}^T_{i,j} [\mathbf{R}_N(\Omega_{1:N})+\gamma^2 \mathbf{I}]^{-1} \boldsymbol{\nu}_{i,j} \}. \label{eq:maxentseq}
\end{split}
\normalsize
\end{align}
\normalsize
\vspace{-0.5cm}
\label{thm:seq}
\end{lemma}

Lemma \ref{thm:seq} provides an \textit{easy-to-evaluate} criterion for greedily maximizing information on $\mathbf{X}$. We give further heuristics for reducing computation time for this optimization in Section \ref{sec:me}.

There are two useful interpretations of the sequential sampling scheme \eqref{eq:maxentseq}. First, it samples the index yielding greatest \textit{information gain} on $\mathbf{X}$, given prior observations $\mathbf{Y}_{\Omega}$. Second, it samples the index with greatest posterior \textit{uncertainty} from the probabilistic model, given observations $\mathbf{Y}_{\Omega}$. This links the objective of information-greedy active learning with the uncertainty quantification of matrix $\mathbf{X}$ (this connection was noted earlier in \citet{mackay1992information}, but in a different active learning context).

\vspace{-0.1in}
\subsection{\texttt{MaxEnt}: Information-theoretic sampling algorithm}
\label{sec:me}
We now combine these insights into an information-theoretic sampling algorithm \texttt{MaxEnt} for low-rank matrix completion. Assume $m_1 \geq m_2$. For \textit{initial} sampling, one way to guarantee a balanced design is to (a) generate $\lfloor m_1/m_2 \rfloor$ random Latin squares \cite{JM1996} of size $m_2 \times m_2$, (b) vertically stacking these squares to form an $(m_2 \lfloor m_1/m_2 \rfloor) \times m_2$ rectangle, and (c) sampling the entries labeled `$a$' from this rectangle. Using these initial samples, the projection matrices $\mathcal{P}_{\mathcal{U}}$ and $\mathcal{P}_{\mathcal{V}}$ can then be estimated via the posterior sampling algorithm \texttt{BayeSMG} in \citet{yuchi2021bayesian} (this is discussed in detail in Appendix \ref{sec:post}). From this, we sample the unobserved entry yielding the greatest expected posterior information gain from \eqref{eq:maxentseq}. These steps are repeated until a desired error is achieved. Algorithm \ref{alg:maxent} summarizes this procedure.

\begin{algorithm}[tb]
\small
  \caption{\texttt{MaxEnt}}
  \label{alg:maxent}
\begin{algorithmic}
  \STATE {\bfseries Input:} Total samples $N_{\max} =N_{ini}+N_{seq}$\\
  \STATE \textit{\underline{Initial sampling}} ($N_{ini} = m_1 \vee m_2$ samples):\\
  $\bullet$ \; Generate \& stack $\lfloor m_1 / m_2 \rfloor$ random $m_2 \times m_2$ Latin squares.\\
  $\bullet$ \; Set $\Omega$ as the entries labeled `a'. 
  \STATE \textit{\underline{Sequential sampling}} ($N_{seq}$ samples): 
  \FOR{$n=N_{ini+1}$ {\bfseries to} $N_{max}$} 
  \STATE $\bullet$ \; Run the posterior sampler \texttt{BayeSMG} \cite{yuchi2021bayesian}.\\
  $\bullet$ \; Obtain projection matrix estimates $(\hat{\mathcal{P}}_{\mathcal{U}},\hat{\mathcal{P}}_{\mathcal{V}})$ via posterior means.\\
  $\bullet$ \; Sample next entry $(i_n,j_n)$ using \eqref{eq:maxentseq}, with $({\mathcal{P}}_{\mathcal{U}},{\mathcal{P}}_{\mathcal{V}})=(\hat{\mathcal{P}}_{\mathcal{U}},\hat{\mathcal{P}}_{\mathcal{V}})$.\\
  $\bullet$ \; Update $\Omega \leftarrow \Omega \cup (i_n,j_n)$.
  \ENDFOR
  \end{algorithmic}
\normalsize
\end{algorithm}

While the sequential sampling procedure makes use of an easy-to-evaluate acquisition function \eqref{eq:maxentseq}, there are several heuristics that can further speed up computation when $\mathbf{X}$ is high-dimensional. First, the exhaustive search in \eqref{eq:maxentseq} over all unobserved indices $\Omega_{1:N}^C$ can be time-consuming. One remedy is to screen out indices with small coherences $\mu_i(\mathcal{U})$ and $\mu_j(\mathcal{V})$ (which are likely poor entries to sample by \eqref{eq:maxentseq}), then perform the search over a much smaller index set. Second, the sequential procedure \eqref{eq:maxentseq} can be extended to a \textit{batch}-sequential procedure, which samples multiple indices simultaneously with large objective values. These heuristics enable \texttt{MaxEnt} to provide efficient information-theoretic sampling for matrices with dimensions on the order of thousands.

\vspace{-0.1in}
\section{Numerical examples}

\subsection{Simulations}

\begin{figure}[!t]
\hfill
\begin{minipage}{0.49\textwidth}
\includegraphics[width=\textwidth]{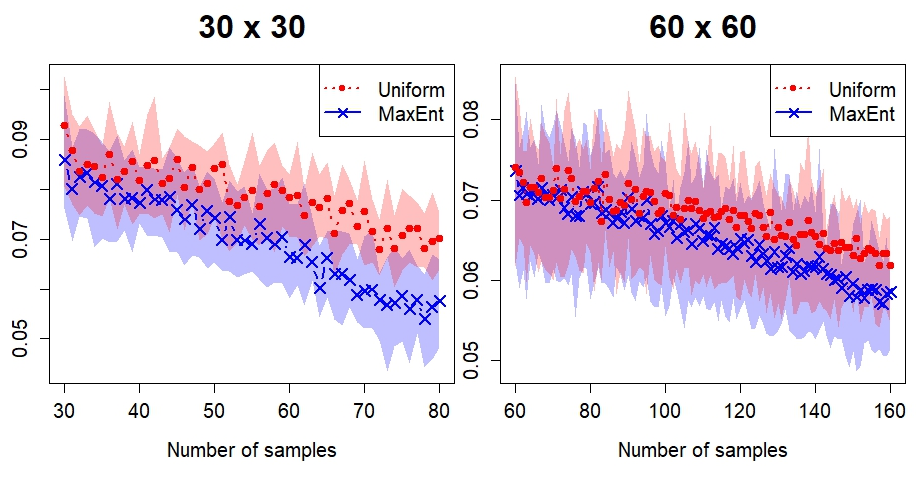}
\vskip -0.1in
\caption{Avg. Frob. errors (line) and 25-th/75-th error quantiles (shaded) for the $30 \times 30$ and $60 \times 60$ simulated matrices.}
\label{fig:seq}
\centering
\includegraphics[width=\textwidth]{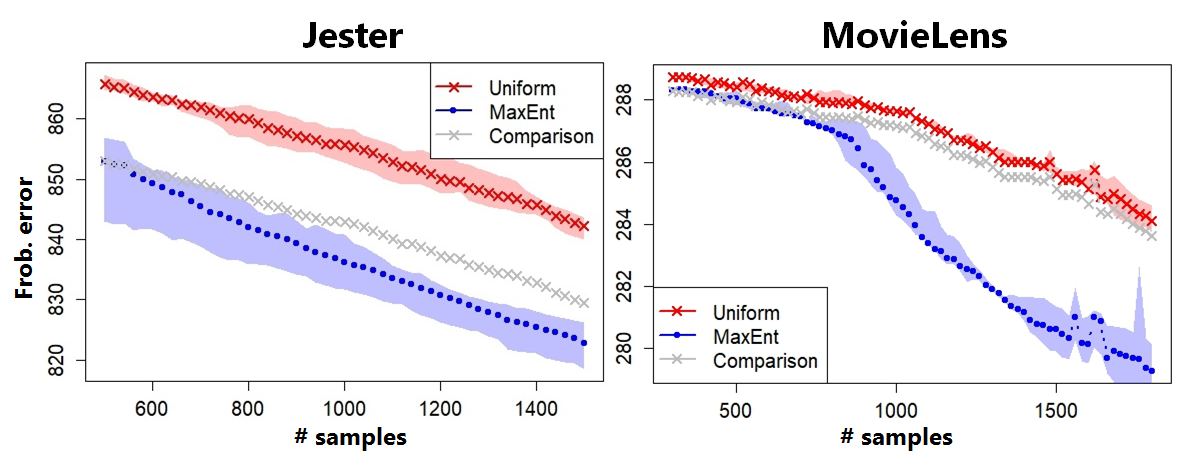}
\vskip -0.1in
\caption{Avg. Frob. errors (line) and 25-th/75-th error quantiles (shaded) for Jester (left) and MovieLens (right).}
\label{fig:jester}
\end{minipage}
\end{figure}

We first investigate the performance of \texttt{MaxEnt} in simulations. The true matrix $\mathbf{X}$ is simulated from the SMG model with priors \eqref{eq:prior}, with hyperparameters $\sigma^2 = 1$, $\eta^2 = 10^{-4}$, $\alpha_{\eta^2} = \alpha_{\sigma^2} = 9$, $\beta_{\eta^2} = 10^{-3}$, $\beta_{\sigma^2} = 10$.
We consider matrices of sizes $30 \times 30$ and $60 \times 60$, with true rank $R = 3$ and $R=4$, respectively. We begin with $N_{ini} = m_1 = m_2$ initial samples, then observe $N_{seq}=50$ and $N_{seq}=100$ entries sequentially for the $30 \times 30$ and $60 \times 60$ cases. This is then replicated 10 times to measure error variability. Figure \ref{fig:seq} shows the averaged errors and error quantiles for \texttt{MaxEnt} and uniform sampling. For initial sampling, \texttt{MaxEnt} yields reduced errors to uniform sampling, which shows the effectiveness of a \textit{balanced} initial design. For sequential sampling, the improvement of \texttt{MaxEnt} over uniform sampling grows larger as more entries are sampled; near the end, the averaged errors from uniform sampling are noticeably higher than the 75\% quantiles from \texttt{MaxEnt}. This shows the effectiveness of our integrated approach, in first (a) learning the underlying subspaces of $\mathbf{X}$ via probabilistic Bayesian modeling, then (b) incorporating this learning to guide information-theoretic sampling.


\subsection{Collaborative filtering}

Next, we apply \texttt{MaxEnt} to two collaborative filtering datasets. The first, `Jester', is collected from the Jester Online Joke Recommender System \cite{Gea2001} -- a test bank of 100 joke ratings from 500 users. 
The second, `MovieLens', is a database of movie ratings for 300 users on 1,700 movies.
For Jester, $N_{ini}=500$ initial samples are collected, with $N_{seq} = 1,000$ entries observed sequentially; for MovieLens, $N_{ini}=300$ and $N_{seq}=1,500$. This is then replicated 10 times to measure error variability. The heuristics from Section \ref{sec:me} are used to speed up the sampling procedure.

Figure \ref{fig:jester} shows the resulting errors using \texttt{MaxEnt} and uniform sampling. Two observations are of interest. First, \texttt{MaxEnt} yields lower initial errors to uniform sampling, which again demonstrates the importance of a balanced initial sample. Second, the gap between \texttt{MaxEnt} and uniform sampling grows larger as entries are observed sequentially, more so than from simulations. One reason is that high coherences are present in both datasets -- there may be users who are overly critical, or jokes or movies which are particularly good. By (a) identifying these preference structures via Bayesian subspace learning, then (b) incorporating this into an information-theoretic sampling procedure, \texttt{MaxEnt} offers an effective way of learning the full rating database from partial observations.
\vspace{-0.1in}
\section{Conclusion}

In this paper, we propose a novel information-theoretic sampling method for noisy matrix completion. Using the Bayesian SMG model \cite{yuchi2021bayesian} as a probabilistic model for the unknown low-rank matrix, we presented an initial and sequential sampling algorithm called \texttt{MaxEnt}, which makes use of subspace learning to guide an information-theoretic sampling of matrix $\mathbf{X}$. Simulations and applications demonstrate the effectiveness of \texttt{MaxEnt} over uniform sampling, and confirm insights developed in the paper. 

\newpage
\bibliography{paper}
\bibliographystyle{icml2021}
\newpage
\appendix
\section{Appendix}
\subsection{\texttt{BayeSMG}: Posterior sampling for UQ}
\label{sec:post}
In this section, we introduce the posterior sampling algorithm for quantifying uncertainty on $\mathbf{X}$ utilized in \ref{sec:me}. 
For efficient sampling, we require a slight reparametrization of $\mathbf{X}$ via its SVD $\mathbf{X} = \mathbf{U}\mathbf{D}\mathbf{V}^T$. Define the \textit{Stiefel manifold} $\mathcal{V}_{R,m}$, the space of $m \times R$ matrices with orthonormal columns (an \textit{$R$-frame} in $\mathbb{R}^m$). By the SVD, the matrix of left and right singular vectors, $\mathbf{U}$ and $\mathbf{V}$, must lie on $\mathcal{V}_{R,m_1}$ and $\mathcal{V}_{R,m_2}$. 
Note that the span of an $R$-frame from the Stiefel manifold $\mathcal{V}_{R,m}$ corresponds to a unique $R$-plane from the Grassmann manifold $\mathcal{G}_{R,m-R}$, but an $R$-plane from $\mathcal{G}_{R,m-R}$ corresponds to infinitely many $R$-frames from $\mathcal{V}_{R,m}$. 
For our model in Section \ref{sec:spec} with the priors configured as in (\ref{eq:prior}), random matrix theory \cite{She2001} then shows: (a) $\mathbf{U}$ and $\mathbf{V}$ are uniformly distributed on $\mathcal{V}_{R,m_1}$ and $\mathcal{V}_{R,m_2}$, and (b) $\mathbf{D} = \text{diag}(\{d_k\}_{k=1}^R)$ follows the so-called \textit{Quadrant Law} (QL; \cite{She2001}).
The uniform distributions on $\mathcal{V}_{R,m_1}$ and $\mathcal{V}_{R,m_2}$ are special cases of the \textit{von Mises-Fisher} (MF) distributions $MF(m_1,R,\mathbf{0})$ and $MF(m_2,R,\mathbf{0})$; a random matrix $\mathbf{W} \sim MF(m,R,\mathbf{F})$ has density \cite{Hof2009}:
\begin{equation}
\small
[\mathbf{W}|R,\mathbf{F}] = \left[ {}_0 F_1 (; \frac{m}{2};\frac{\mathbf{F}^T\mathbf{F}}{4}) \right]^{-1} \text{etr}(\mathbf{F}^T \mathbf{W}), \; \mathbf{W} \in \mathcal{V}_{R,m},
\normalsize
\label{eq:mf}
\end{equation}
where ${}_0 F_1 (;\cdot;\cdot)$ is the hypergeometric function. The singular values $\mathbf{D}$ follow $QL(\mathbf{0},\sigma^2)$, where $QL(\boldsymbol{\mu},\delta^2)$ is the quadrant law with density:
\begin{equation}
\small
[\mathbf{D}|\boldsymbol{\mu},\delta^2] = \frac{\exp\left\{ -\frac{1}{2 \delta^2} \sum_{k=1}^R (d_k - \mu_k)^2 \right\}}{Z_R(2\pi\delta^2)^{R/2}} \prod_{\substack{k,l=1; k < l}}^R |d_k^2 - d_l^2|,
\normalsize
\label{eq:ql}
\end{equation}
and $Z_R$ is a normalization constant depending on $R$.
\begin{algorithm}[H]
    \small
  \caption{\texttt{BayeSMG}}
  \label{alg:gibbs}
\begin{algorithmic}
  \STATE {\bfseries Input:} $\mathbf{Y}_{\Omega}$, $R$, $\alpha_{\eta^2}, \beta_{\eta^2}, \alpha_{\sigma^2}, \beta_{\sigma^2}$,
  \STATE Complete $\mathbf{X}_0$ from $\mathbf{Y}_{\Omega}$ via \eqref{eq:nuc}.
  \STATE Initialize $[\mathbf{U}_0,\mathbf{D}_0, \mathbf{V}_0] \leftarrow \text{svd}(\mathbf{X}_0)$, $\eta^2_0$ and $\sigma^2_0$.
  \STATE \textit{\underline{Gibbs sampler}}: 
  \FOR{$t=1$ {\bfseries to} $T$}
  \STATE $\mathbf{X}_t \leftarrow \mathbf{U}_{t-1}\mathbf{D}_{t-1}\mathbf{V}_{t-1}^T$.
  \STATE $\mathbf{Y}_{\Omega^c} \sim \mathcal{N}(\mathbf{X}^P_{\Omega^c},\boldsymbol{\Sigma}^P_{\Omega^c}+\eta^2 \mathbf{I})$.
  \STATE $\mathbf{U}_t \sim MF(m_1,R,{\mathbf{Y} \mathbf{V}_{t-1} \mathbf{D}_{t-1}}/{\eta^2_{t-1}})$.
  \STATE $\mathbf{V}_t \sim MF(m_2,R,\mathbf{Y}^T \mathbf{U}_t \mathbf{D}_{t-1}/\eta^2_{t-1})$.
  \STATE $\mathbf{D}_t \sim QL(\boldsymbol{\mu},\delta^2)$, 
  \STATE where $\boldsymbol{\mu} = [\sigma^2_{t-1} \mathbf{u}_{k,t}^T \mathbf{Y} \mathbf{v}_{k,t} / (\eta^2_{t-1} + \sigma^2_{t-1})]_{k=1}^R$
  \STATE $\delta^2 = \eta^2_{t-1} \sigma^2_{t-1} / (\eta^2_{t-1} + \sigma^2_{t-1})$.
\item $\sigma^2_t \sim IG( \alpha_{\sigma^2} + R/ 2, \beta_{\sigma^2} + \text{tr}(\mathbf{D}_t^2) / 2 )$
\item $\eta^2_t \sim IG( \alpha_{\eta^2} + m_1 m_2 / 2, \beta_{\eta^2} + \|\mathbf{Y} - \mathbf{X}_t\|_F^2/2)$.
  \ENDFOR
  \STATE Return posterior samples $\{(\mathbf{X}_t, \mathbf{U}_t, \mathbf{V}_t)\}_{t=1}^T$.
\end{algorithmic}
\normalsize
\end{algorithm}





\end{document}